\documentclass{article}

\PassOptionsToPackage{numbers, compress}{natbib}

\usepackage[preprint]{neurips_2025}

\usepackage[utf8]{inputenc} 
\usepackage[T1]{fontenc}    
\usepackage[hidelinks, colorlinks=true,
    linkcolor=black,
    urlcolor=blue,
    citecolor=black]{hyperref}       
\usepackage{url}            
\usepackage{booktabs}       
\usepackage{amsfonts}       
\usepackage{amsmath}        
\usepackage{nicefrac}       
\usepackage{microtype}      
\usepackage{xcolor}         
\usepackage{graphicx}       
\usepackage{comment}
\usepackage{wrapfig}        
\usepackage{svg}
\usepackage[hang,flushmargin]{footmisc}
\usepackage[dvipsnames]{xcolor}

\usepackage{graphics}

\makeatletter
  \def\title@font{\Large\bfseries}
  \let\ltx@maketitle\@maketitle
  \def\@maketitle{\bgroup%
    \let\ltx@title\@title%
    \def\@title{\resizebox{\textwidth}{!}{%
      \mbox{\title@font\ltx@title}%
    }}%
    \ltx@maketitle%
  \egroup}
\makeatother

\newcommand{\corpus}{\textsc{SQaLe}}

\title{\textsc{SQaLe}: A Large Text-to-SQL Corpus Grounded in Real Schemas}

\workshoptitle{AI for Tabular Data}

\author{%
  \textbf{Cornelius Wolff}$^{1,2}$ \quad Daniel Gomm$^{1,2}$ \quad Madelon Hulsebos$^1$ \\
  $^1$Centrum Wiskunde \& Informatica \quad $^2$University of Amsterdam\\
  \texttt{\{cornelius.wolff, daniel.gomm, madelon.hulsebos\}@cwi.nl} \\
}

\makeatletter
\renewcommand{\@noticestring}{AI for Tabular Data workshop at EurIPS 2025}
\makeatother

\begin{document}

\maketitle

\begin{abstract}
Advances in large language models have accelerated progress in text-to-SQL, methods for converting natural language queries into valid SQL queries. A key bottleneck for developing generalizable text-to-SQL models is the lack of large-scale datasets with sufficient schema and query complexity, domain coverage, and task diversity. We introduce \corpus{}, a large-scale semi-synthetic text-to-SQL dataset built on 135{,}875 relational database schemas expanded from a collection of real-world schemas, SchemaPile. We establish a principled generation pipeline which combines schema sampling, question synthesis, and SQL construction, and produce 517{,}676 high-quality (question, schema, query) triples. The \corpus{} dataset captures realistic schema size variability, diverse query patterns, and natural language ambiguity while maintaining execution validity. We provide an analysis of its contents and characteristics, and find that \corpus{} introduces the most realistic large-scale text-to-SQL dataset to date in comparison with existing benchmarks and datasets. We discuss how \corpus{} enables our vision for data scaling and model generalization in text-to-SQL research. The dataset is accessible at: \href{https://huggingface.co/datasets/trl-lab/SQaLe-text-to-SQL-dataset}{https://huggingface.co/datasets/trl-lab/SQaLe-text-to-SQL-dataset}.
\end{abstract}

\section{Introduction}

When it comes to the retrieval of insight from large document stores, large language models (LLMs) and retrieval-augmented generation (RAG) have significantly advanced natural language understanding and question answering by reducing bottlenecks in unstructured document retrieval \cite{lewis_retrieval-augmented_2021, gao_retrieval-augmented_2024}. However, when it comes to structured data, the development of dedicated models for accessing structured data through text-to-SQL remains constrained by the limited availability of large, diverse, and high-quality datasets \cite{liu_survey_2025}. Although LLMs have enabled rapid progress in text-to-SQL generation, building new and generalizable models from scratch still faces major challenges due to insufficient data scale and schema diversity \cite{gao_text--sql_2024}. This problem is fundamental to enabling natural language interfaces for databases.

Over the past years, the field has evolved through successive generations of datasets that attempt to balance realism, schema complexity, and linguistic variety. The first common benchmark was \emph{Spider~1.0}~\cite{yu_spider_2018}, which introduced cross-database generalization and multi-table reasoning, containing around 10{,}000 questions across hundreds of databases. It quickly became the de facto standard for Text-to-SQL evaluation. It was later largely replaced by \emph{BIRD}~\cite{li_can_2023}, which placed greater emphasis on realism by incorporating practical database contexts. Other corpora such as \emph{KaggleDBQA}~\cite{lee_kaggledbqa_2021} contributed similar aims, while domain-specific datasets including \emph{EHRSQL}~\cite{lee_ehrsql_nodate} and others~\cite{wang_text--sql_2020} targeted specialized applications. More recently, \emph{Spider~2.0}~\cite{lei_spider_2025} has extended this line of work to enterprise settings, featuring more diverse schemas and queries, and is rapidly becoming a new common benchmark. Despite these advances, the overall scale of available resources remains orders of magnitude smaller than standard NLP corpora~\cite{liu_datasets_2024}, making them insufficient under established scaling laws for training large models~\cite{kaplan_scaling_2020, hoffmann_training_nodate}. While synthetic and semi-synthetic approaches have begun to address these gaps, only a few datasets, such as \emph{SynSQL-2.5M}~\cite{li_omnisql_2025}, have reached million-scale coverage, relying primarily on artificial schemas and programmatic question generation. Yet, the used schemas often lack the scale, complexity, and other real-world characteristics needed to fully capture the challenges of practical Text-to-SQL tasks~\cite{gao_text--sql_2024, yang_synthesizing_2024, liu_survey_2025}.

To address the limitations of scale, diversity, and realism, we construct a large-scale Text-to-SQL dataset with broad schema coverage and naturally phrased questions. The \corpus{} dataset contains 135{,}875 relational database schemas extracted from SchemaPile’s 22{,}989 real-world database schemas~\cite{dohmen_schemapile_2024} and contains 517{,}676 (question, schema, query) triples. Unlike fully synthetic resources, our generation pipeline is grounded in real schemas and guided by principled linguistic and structural criteria to ensure both realism and diversity in queries, resulting in distinctive characteristics such as . We also provide a descriptive analysis of \corpus{} in comparison with existing datasets, and discuss its implications for training generalizable text-to-SQL models. Our contributions are as follows: \textbf{(1)} a data generation pipeline for synthesizing text-to-SQL data from real schemas; \textbf{(2)} the creation of a large-scale text-to-SQL dataset; and \textbf{(3)} an in-depth characterization of the dataset, highlighting its scalability potential and relevance for model training and evaluation.

\section{Creating Text-to-SQL Data at Scale}

Here, we describe our methodology for producing \corpus{}: a principled, scalable pipeline that synthesizes representative (schema, question, query) triples corresponding with real database schemas.

\subsection{Dataset Design Criteria}

The design of large-scale text-to-SQL datasets should follow principles that ensure both representational realism and linguistic-semantic diversity. We distinguish between schema-level and query-level considerations. These criteria are grounded in empirical analyses of production databases and informed by benchmarks such as Spider~2.0~\cite{lei_spider_2025}, BIRD~\cite{li_can_2023}, and SQLStorm~\cite{schmidt_sqlstorm_2025}. All of these criteria were furthermore confirmed through expert interviews with data engineers and practitioners.

\subsubsection{Schema-Level Criteria}

\textbf{Schema Size (\textbf{C1}).}  
A realistic dataset should include schemas ranging from small, domain-specific databases to large, enterprise-scale systems containing hundreds of tables. This variety reflects the scale of modern data ecosystems observed in industry and academic corpora~\cite{cleve_understanding_2015,cortez_annotating_2015,lei_spider_2025}.

\textbf{Schema Density and Normalization (\textbf{C2}).}  Text-to-SQL datasets should capture variation in schema density, the average number of columns per table, as a proxy for normalization depth and granularity, as observed in real-world databases \cite{dohmen_schemapile_2024,brahmia_literature_2024}. Furthermore, including both highly normalized and denormalized schemas reflects different relational schema principles\cite{kohita_exploring_2025}.

\textbf{Foreign Key Integrity (\textbf{C3}).}  
Datasets should represent realistic variability in referential integrity. Many real-world databases contain incomplete or implicit foreign key relationships; including such cases ensures that models are exposed to typical challenges of schema reasoning and foreign-key inference. This aspect of real-world databases has been a research topic for years and therefore should be reflected in the dataset \cite{rostin_machine_nodate, zhang_multi-column_2010, katsogiannis-meimarakis_-depth_2026}.

\textbf{Naming Conventions (\textbf{C4}).}  
Schema naming should reflect authentic, heterogeneous practices, including inconsistent abbreviations, mixed casing, and domain specific terms. Preserving this variability enhances validity and prevents overfitting to overly sanitized or idealized schemas~\cite{dohmen_schemapile_2024,lei_spider_2025}.

\subsubsection{Query-Level Criteria}

\textbf{Join Complexity (\textbf{C5}).}  
A well-designed text-to-SQL dataset should span a distribution of SQL join complexities, from single-table to multi-join analytical queries. This range allows for balanced evaluation of compositional reasoning and structural generalization~\cite{lei_spider_2025,schmidt_sqlstorm_2025,chen_beaver_2025}.

\textbf{Operator Diversity (\textbf{C6}).}  
Queries should include diverse SQL constructs such as aggregations, comparisons, nested subqueries, set operations, and logical combinations to represent the breadth of real analytical workloads and support model training on varied syntactic and semantic forms \cite{liu_survey_2025, lan_unite_2023}.

\textbf{Intent Diversity (\textbf{C7}).}  
Natural language inputs should encompass a wide spectrum of user intents, such as lookup, aggregation, filtering, comparison, and ranking. Coverage of distinct intent categories enables more comprehensive assessment of a model’s semantic understanding capabilities \cite{dragusin_grounding_nodate, lan_unite_2023}.

\textbf{Query Ambiguity (\textbf{C8}).}  
High-quality datasets should contain controlled linguistic and semantic ambiguity, reflecting the uncertainty that stems from a user's incomplete understanding of the underlying data, such as unfamiliarity with the schema or the specific values of interest \cite{bhaskar_benchmarking_2023, floratou_nl2sql_2024, wang_handling_nodate}.

\subsection{The \corpus{} Generation Pipeline}

\begin{figure}[t!]
    \centering
    \includegraphics[width=\linewidth]{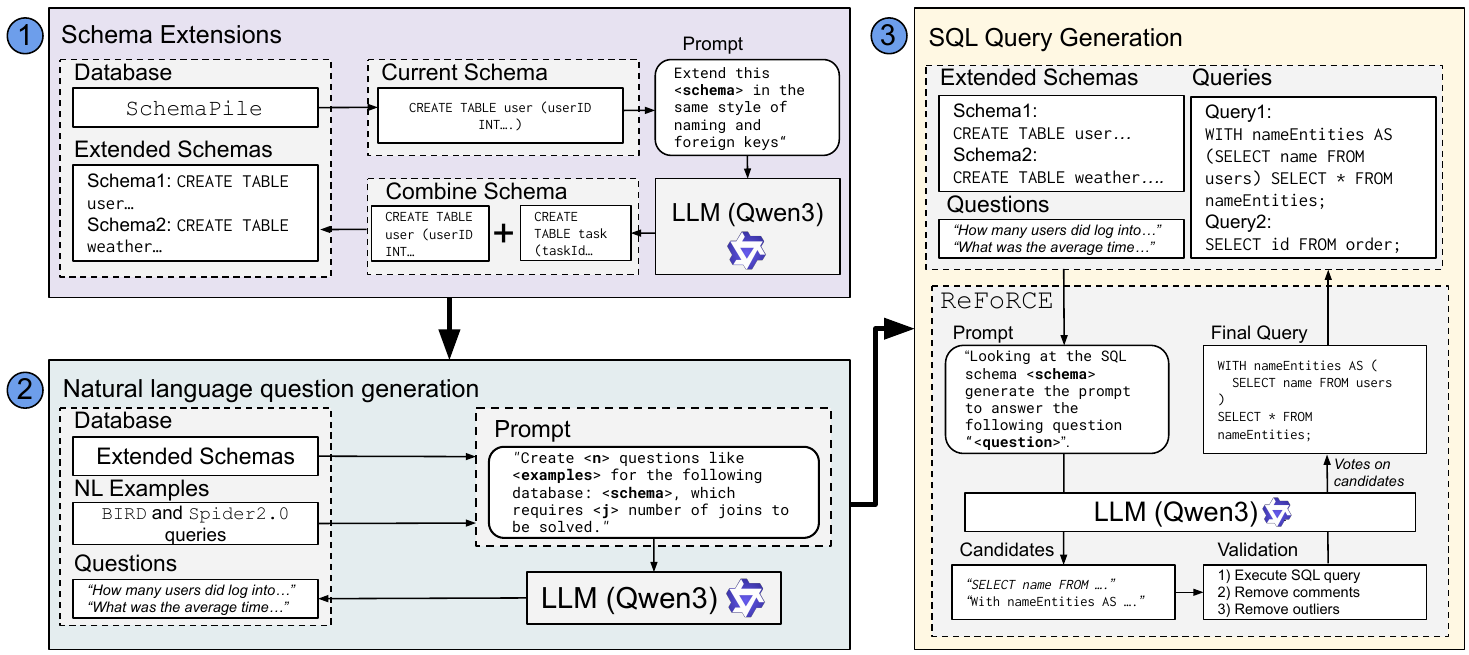}
    \vspace{-0.2cm}
    \caption{Overview of the data creation pipeline for the \corpus{} dataset. The displayed prompts are simplified for clarity.}
    \label{fig:creation-pipeline}
\end{figure}

The \corpus{} generation pipeline (Figure \ref{fig:creation-pipeline}) follows a structured, multi-stage approach that integrates schema extension, question synthesis, and SQL generation. It begins with a repository of real-world database schemas, SchemaPile \citep{dohmen_schemapile_2024}, which serves as the foundation for realistic structural diversity (\textbf{C2}). Using a large language model (Qwen3 30B \citep{yang_qwen3_2025}), existing schemas are \textbf{extended incrementally}, for example by including 5 to 25 synthetic tables, until the target query complexity distribution is balanced around a median of 4 joins per query, informed by SQLStorm metrics \cite{schmidt_sqlstorm_2025} (\textbf{C1}). The targeted schema size, in number of tables, follows a smooth gamma-like probability distribution, where the mode is set to 100 and the maximum value to 350 based on the maximum in Spider 2.0 \cite{lei_spider_2025}. During the extension process, the LLM is prompted to maintain style, naming conventions and level of normalization of the original schema from SchemaPile (\textbf{C2, C4}). All \textbf{intermediate schema sizes} generated during the iterative extension process are also retained in the dataset, ensuring balanced representation across schema scales. Furthermore, for all schemas inferred from schemas without foreign key relations, any foreign keys added during extension process are removed (\textbf{C3}).

In the next stages, the pipeline performs \textbf{question generation} and \textbf{SQL query creation} using the same LLM with guided prompting. For each extended schema, natural language questions are synthesized from examples drawn from BIRD and Spider2.0, ensuring linguistic diversity, while questions are selected in line with the targeted join distribution (\textbf{C6–C8}). We chose this open question generation over a completely template based approach like in \cite{papicchio2023qatch} in order to maximize the variety of questions in terms of style, length and complexity. Example selection also balances easy and difficult joins to ensure diversity (\textbf{C5}). These questions are then paired with candidate SQL statements in a \textbf{query creation phase} using the ReFoRCE text-to-SQL framework~\citep{deng_reforce_2025}, which generates multiple SQL candidates, applies a voting process to select the best one, and validates executability by executing the SQL statement against the schema and filter for outliers in terms of prompt and query length. Finally, we manually randomly checked 300 samples from the dataset to ensure quality and correctness.

\section{The \corpus{} Dataset}

\begin{table}[h!]
\centering
\caption{Schema statistics for datasets. M. stands for median.}
\label{tab:schema_summary}
\vspace{-0.1cm}
\begin{tabular}{lrrrrr}
\toprule
\textbf{Dataset} & \textbf{\#Schemas} & \textbf{M. \#cols/schema} & \textbf{M. \#tables/schema} & \textbf{\#FKs} \\
\midrule
BIRD (train \& dev) & 80 & 39 & 5.0 & 526 \\
EHRSQL & 2 & 92 & 13.5 & 34 \\
Spider2 & 236 & 89 & 7.0 & 0 \\
SynSQL & 16{,}575 & 72 & 10.0 & 159{,}547 \\ \midrule
\corpus{} & 135{,}875 & 435.0 & 91.0 & 13{,}201{,}052 \\
\bottomrule
\end{tabular}
\end{table}

\begin{table}[h!]
\centering
\caption[]{Query statistics for datasets. Where and Join columns indicate the share of queries using those operators and the Func column contains multiple operators\footnotemark. M. indicates the median.}
\label{tab:dataset_summary_no_select}
\vspace{-0.1cm}
\resizebox{\textwidth}{!}{
\begin{tabular}{lrrrrrrr}
\toprule
\textbf{Dataset} & \textbf{\#queries} & \textbf{M. \#tables/query} & \textbf{M. \#tokens/query} & \textbf{Func. (\%)} & \textbf{Where (\%)} & \textbf{Join (\%)} \\
\midrule
BIRD (train \& dev) & 10{,}962 & 2 & 41 & 13.8 & 88.1 & 76.2 \\
EHRSQL & 9{,}270 & 2 & 83 & 76.9 & 99.9 & 19.7 \\
Spider2 & 250 & 3 & 229.5 & 45.2 & 94.4 & 72.0 \\
SynSQL & 2{},544{,}390 & 3 & 85 & 12.1 & 75.6 & 89.4 \\ \midrule
\corpus{} & 517{,}676 & 3 & 61 & 26.9 & 78.9 & 76.1 \\
\bottomrule
\end{tabular}
}
\vspace{-0.5cm}
\end{table}

Using the outlined generation pipeline, the \corpus{} dataset represents a large-scale, semi-synthetic Text-to-SQL resource grounded in real-world database structures. The generation pipeline, executed with up to 100x H100 GPUs in parallel, results in 517{,}676 validated (schema, question, query) triples.

\begin{figure}[h!]
    \centering
    \begin{minipage}{0.325\linewidth}
        \centering
        \includegraphics[width=\linewidth]{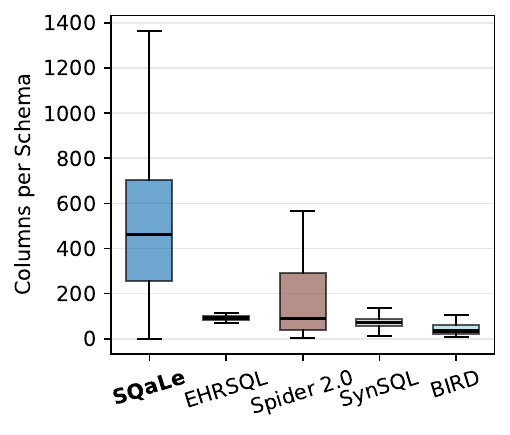}
        
    \end{minipage}
    \hfill
    \begin{minipage}{0.325\linewidth}
        \centering
        \includegraphics[width=\linewidth]{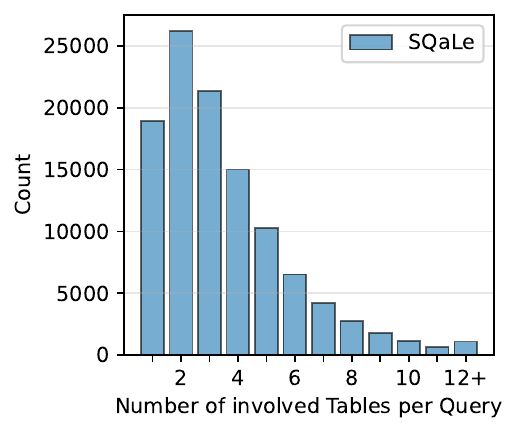}
    \end{minipage}
    \hfill
    \begin{minipage}{0.325\linewidth}
        \centering
        \includegraphics[width=\linewidth]{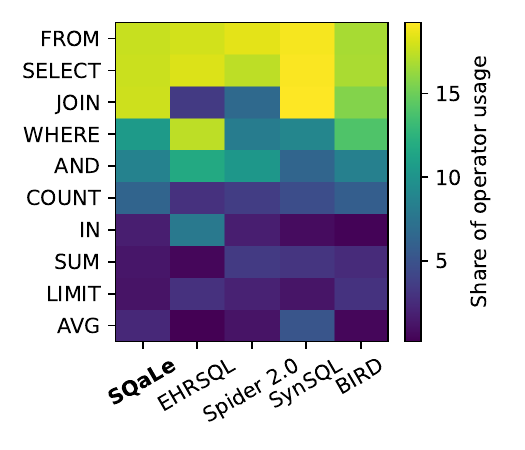}
    \end{minipage}
    \caption{\corpus{} data statistics. Left: distribution of column counts. Center: Distribution of the number of tables involved in queries. Right: Share of specific operators among all operators.}
    \label{fig:column-counts}
\end{figure}

Compared to existing datasets, \corpus{} offers a substantially broader and more realistic schema distribution (Table \ref{tab:schema_summary} and Figure \ref{fig:column-counts}). With 135{,}875 schemas, a median of 91 tables and 435 columns/schema, and over 13 million Foreign Keys (FK) relations, it far exceeds the structural diversity of Spider 2.0, SynSQL and BIRD. While not as large as SynSQL in terms of total number of queries, our dataset features vastly more and more complex schemas, emphasizing depth and structural realism over volume. In terms of query composition (Table \ref{tab:dataset_summary_no_select} and Figure \ref{fig:column-counts}), \corpus{} shows rich operator diversity and join complexity, with 76.1\% involving joins. These proportions closely match the complexity of BIRD and Spider 2.0 while extending further into multi-join and nested queries.

The scale of \corpus{} is deliberately chosen to support the development of new small-scale text-to-SQL models and fine-tuning of larger foundation models. Grounded in authentic schema patterns and validated through the rigorous ReFoRCE-based SQL validation process, the dataset balances realism and efficiency, providing high-quality data for building and adapting text-to-SQL models. It addresses current inefficiencies seen in leaderboards~\citep{lei_spider_2025, li_can_2023}, where large models and complex pipelines dominate, and even ReFoRCE requires 12 seconds on an H100 to generate a single SQL query. The dataset can be accessed at \href{https://huggingface.co/datasets/trl-lab/SQaLe-text-to-SQL-dataset}{https://huggingface.co/datasets/trl-lab/SQaLe-text-to-SQL-dataset}.

\paragraph{Toward scaling text-to-SQL with \corpus{}.} Scaling laws in machine learning suggest that larger datasets yield larger models that will continue to perform better~\cite{kaplan_scaling_2020, hoffmann_training_nodate}. Informed by task-specific scaling studies, we hypothesize that through significantly larger, more complex and diverse text-to-SQL datasets we can achieve proportional gains in model performance as a function of dataset scale. To explore the effectiveness of large-scale pretraining of text-to-SQL models, we introduce the text-to-SQL dataset \corpus{} consisting of 517{,}676 semi-synthetic (schema,query,SQL) triples grounded in real-world schemas and queries. A sample of the dataset can be seen in appendix \ref{sec:dataset_sample}.

\footnotetext{CAST, COALESCE, SUBSTR, LENGTH, UPPER, LOWER, ROUND, ABS, DATE, DATETIME, STRFTIME.}

\section*{Acknowledgments} \vspace{-0.2cm}
This work was partially funded by grants from NWO (NGF.1607.22.045) and SAP.

{\hypersetup{urlcolor=black, linkcolor=black, citecolor=black}
\bibliographystyle{plainnat}
\bibliography{references}


\newpage
\appendix

\section{Dataset Sample}
Below are two representative entries from the \corpus{} dataset illustrating the DDL of real schemas, the corresponding natural-language questions, the validated SQL queries, and concise metadata (token counts, join counts, involved tables, and columns) like it can be found on HuggingFace.
\label{sec:dataset_sample}
\begin{table}[ht]
\centering
\small
\resizebox{\columnwidth}{!}{%
\begin{tabular}{|p{3cm}|p{2.5cm}|p{3cm}|p{1.8cm}|l|l|l|l|}
\hline
\textbf{Schema (DDL)} & \textbf{Question} & \textbf{Query} & \textbf{Token Count} & \textbf{Joins} & \textbf{Tables} & \textbf{Cols} \\ \hline
\texttt{CREATE TABLE courses (course\_id TEXT, name TEXT, teacher\_id TEXT...} &
List all tasks with course names and task states. &
\texttt{SELECT tasks.name, courses.name FROM tasks JOIN courses ON ...;} &
\{q:57, s:338, tot:505\} &
3 &
6 &
25 \\ \hline
\texttt{CREATE TABLE employees (id INT, name TEXT, dept TEXT, salary ...} &
Find total salary by department. &
\texttt{SELECT dept, SUM(salary) FROM employees GROUP BY dept;} &
\{q:22, s:120, tot:142\} &
0 &
2 &
5 \\ \hline

\end{tabular}
}
\caption{Example entries illustrating the structure of the SQL generation dataset.}
\label{tab:dataset_structure}
\end{table}


\end{document}